# Neutron Scattering Instrumentation at Compact Neutron Sources

Monday July 17th – Tuesday July 18th 2018
Centre de Formation du CNRS, Gif sur Yvette

*Sarah Böhm[1], Tobias Cronert[2], Jan-Philipp Dabruck[1], Xavier Fabrèges[3], Thomas Gutberlet[2], F. Mezei[4], Alain Letourneau[5], Alain Menelle[3], Frédéric Ott[3], Florence Porcher[3], Ulrich Rücker[3], Hoang Tran[3], Jörg Voigt[2], Paul Zakalek[2]*

[1] RWTH Aachen Universität, Institut für Nukleare Entsorgung und Techniktransfer, Wüllnertrsaße 2, 52062 Aachen

[2] Jülich Centre for Neutron Science JCNS, Forschungszentrum Jülich, Lichtenbergstrasse 1, 85747 Garching

[3] Laboratoire Léon Brillouin, UMR12, CEA, CNRS, Université Paris-Saclay, CEA Saclay, 91191 Gif sur Yvette.

[4] European Spallation Source, ESS, ERIC Box 176, SE-221 00 Lund

[5] Département de Physique Nucléaire, Université Paris-Saclay, CEA Saclay, 91191 Gif sur Yvette.

## Abstract

There is currently a strong interest in Compact Accelerator-based Neutron Source (CANS) as a possible new type of source for neutron scattering experiments. A workshop around the "Neutron scattering instrumentation around CANS" was organized in July 2017 between several European institutes. This report summarizes the main outcome of the discussions.

The document is aiming at providing general guidelines for the instrumentation around CANS. Detailed technical discussions are or will be provided in specific publications.

## Keywords:

- Compact Accelerator-based Neutron Source
- CANS
- Neutron Scattering Instrumentation

## Corresponding author

Frédéric OTT    Frederic.Ott@cea.fr

# Introduction

As soon as neutron reactors were operating they have been used for neutron scattering experiments. Once the potential of neutron scattering was unraveled for magnetic studies and spectroscopic measurements, dedicated facilities were built. The still world leading facility, the Institut Laue Langevin, started operation as early as 1971. It is still operating the most performing instruments nowadays. There are currently about 50 nuclear research reactors operating across the world who are performing neutron scattering experiments [1]. Among these facilities, 20 are running a user program, that is, they are offering the possibility to academic users to perform neutron scattering experiments. During the 1980', a new type of neutron facilities based on the spallation reaction was developed. Eventually this led to the creation of 4 new facilities (KENS in Japan, IPNS [2], Los Alamos Neutron Science Center [3] and ISIS [4]) which were also running a user program. During the 2000', a second generation of spallation sources was built (SNS in the USA [5] and JPARC in Japan [6]). ESS [7] which can be considered as a third generation spallation source is currently being built in Europe.

These last sources are very powerful and are able to replace or overtake nuclear reactors in terms of performances for neutron scattering. Considering the situation in Europe regarding licensing and operational requirements, it is very unlikely that new nuclear research reactors will be considered to be built in replacement of the old ones. However, currently aging facilities are providing a broad and engaged user base feeding constant high interest in the most performing ones. It is thus necessary to look for solutions to ensure that this user base (6000 users [8]) can be maintained in Europe in order to secure the scientific championship and make the best use of the most powerful sources. The price tag of a full fledge spallation source is quite high (~1B€) and is difficult to bare by a single country. Hence possibilities to build neutron scattering facilities which could replace existing nuclear reactors is very welcome provided the investment would make it affordable to a single country on par with a comparable synchrotron or a power laser facility.

During the last decade a number of groups have independently considered the possibility of operating a high current / low energy proton accelerator to produce thermal and cold neutrons [9-10-11-12-13-14]. A few facilities have actually been built and are operating scattering instruments [15]. A UCANS network gathering these groups has been created [16]. The core idea of these projects is that low energy accelerator sources, if properly optimized for a well identified task, can provide a neutron flux suitable for a number of experiments and research topics.

A workshop around the "Neutron scattering instrumentation around CANS" was organized in July 2017 between several European institutes. This report summarizes the main outcome of the discussions. Several aspects were discussed, the state of the art at existing CANS, the expected performances of various types of neutron scattering instruments on various CANS designs, the moderator optimization issues and the way such facilities could be operated compared to the existing situation.

# State of the art

Accelerator based neutron sources do exist since several decades but they have been mostly based on photo-fission neutrons produced with electron beams. This is historically due to the fact that electron accelerators are significantly easier to build than proton accelerators. The Harwell facility used an electron linac to produce neutrons with a uranium target as early as 1967 and performed neutron scattering experiments [17]. Facilities such as HUNS in Japan or the Bariloche LINAC in Argentina also started operation during the 1970'. While the HUNS facility is still in operation and has been upgraded recently, the facility in Bariloche has stopped operation in 2017. These facilities have mostly focused on nuclear physics experiments and education.

During the last 20 years significant developments have been made in the field of accelerators and the possibility of operating accelerators with ion currents on the order of 100mA has been demonstrated. About 15 years ago this has led to an interest in setting up neutron sources based on proton accelerators for neutron scattering. Four documented facilities are described below. In 2009, the UCANS Union for Compact Accelerator Neutron Source has been created (http://www.ucans.org/).

## The LENS compact neutron source

The LENS facility is based at the Bloomington University Indiana. The nominal parameters are 13MeV, $I_{peak}$ = 20mA, duty cycle ~1%, $P_{max}$ = 2.6 kW. It is operating a SANS instruments, a Spin-Echo instrument, and a radiography instrument. Some SANS data show that it is possible to measure SANS signal down to 0.1cm$^{-1}$ (Das et al, Langmuir 2014) which is only 10 times higher than measurements at regular SANS machines on existing user facilities.

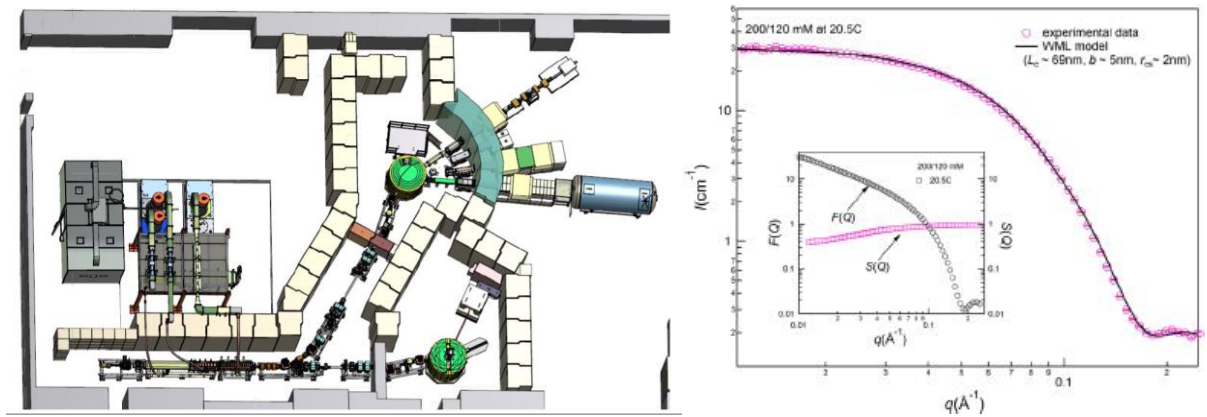

*Figure 1: (left) LENS layout with 2 target stations (green) and 3 instruments around the 2nd target station. (right) SANS data. CTAB (200mM) micelles with 120 mM NaCl. Measured at LENS@13MeV; 20mA; 20Hz, 600µs; $I_{av}$ = 0.24mA ; P = 3kW (Das et al, Langmuir 2014, [18]).*

## The RANS compact neutron source

The RANS source at RIKEN in Japan does not yet operate dedicated instruments but performs experiments "on demand". Simple instruments are set-up around the source as needed. The source operates at $E_p$ = 7 MeV ; $I_{av}$ = 100µA ; $P_{max}$ = 700W.

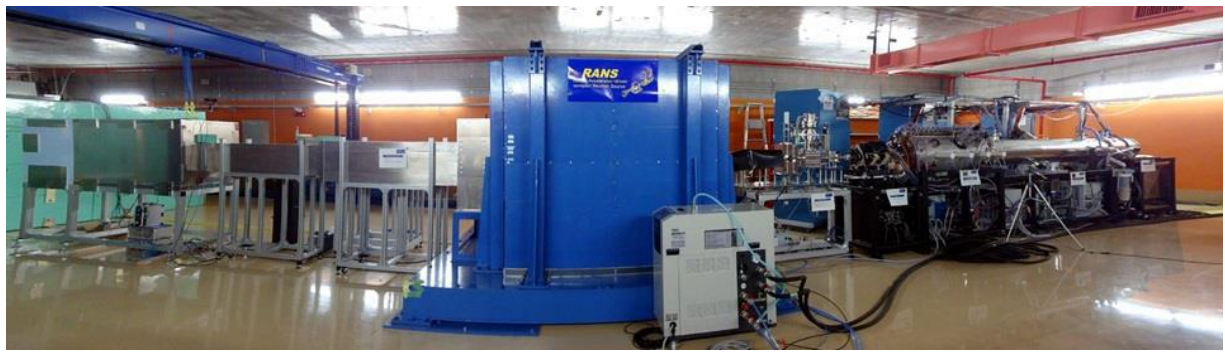

*Figure 2: The RANS casemate. (left) scattering instrument (middle) TMR assembly, mostly shielding, (right) accelerator.*

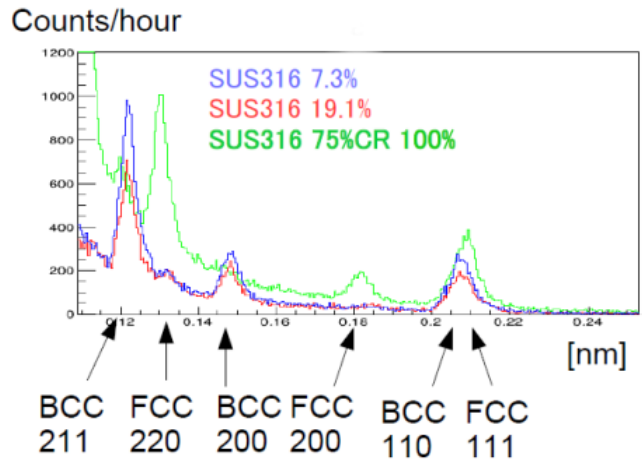

*Figure 3: Powder diffraction patterns on steel samples made to obtain austenite – martensite ratios.*

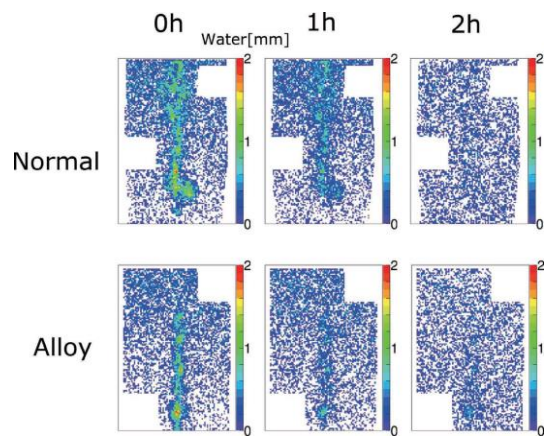

*Figure 4: Radiography of corroded steel plates and humidity up-take as a function of time. Pixel Size 0.8x0.8mm² ; 5 minutes exposure time; $E_p$ = 7 MeV ; $I_{av}$ = 15µA ; P = 100W.*

## The HUNS source (Hokkaido University)

Hokkaido University neutron source, HUNS was completed in 1973, and has been used actively for developments of moderators, neutron instruments, neutron devices and new methods for 40 years although its power is not so high. Recently, a pulsed neutron imaging method has been developed and a new type of small angle neutron scattering method has been also developed.

The source design parameters are $E_{electrons}$ = 35-45 MeV ; P = 1kW ; 20K methane moderator.

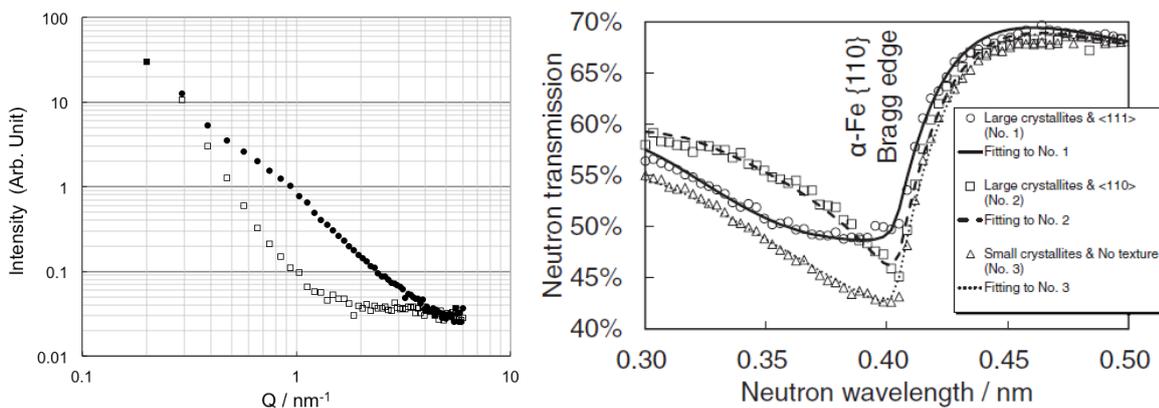

*Figure 5: (left) SANS in steel samples with (filled markers) and without (open markers) nanoscopic precipitates. (right) Bragg-edge transmission spectra measured at HUNS, and the profile fitting curves obtained by RITS. [19]*

## The CPHS source

China is putting huge efforts in neutron scattering: the 60 MW CARR reactor, the China Spallation Neutron Source (CSNS), the 20 MW China Mianyang Research Reactor (CMRR) are all designed to be user facilities. Besides, the CPHS is a less ambitious project of a compact source based in Beijing. The parameters of the source are $E_p$ = 13MeV, $I_{peak}$ = 50mA, duty cycle = 2.5%, P = 16.3kW. It is however not yet operating at the nominal proton energy of 13 MeV but is limited at $E_p$ = 3MeV.

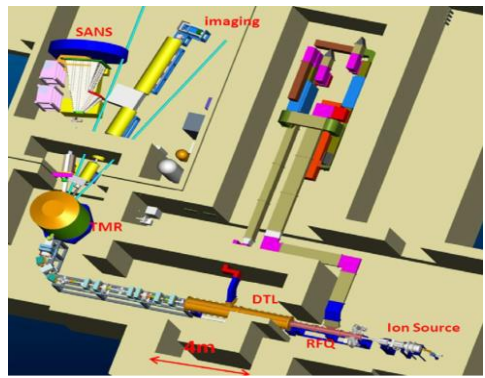

*Figure 6: Layout of the CPHS facility*

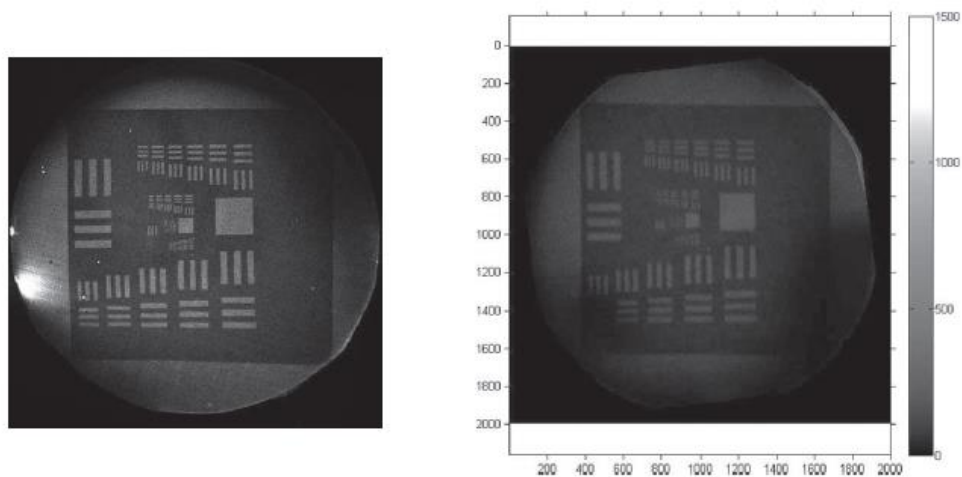

*Figure 7: the MCP image of a USAF-1951 Gd-mask measured with the beam line of CPHS (left) and CARR (right). Note that the measuring conditions are not documented (measuring time, L/D ratio, CARR power...)*

## Neutron scattering instrumentation around CANS

In Europe several institutes are considering high-end CANS facilities using the latest available technologies to achieve High Brilliance using low energy accelerators. Instrument performances on instruments installed on an optimized High Brilliance Source (HBS) will be optimized to match the performances of comparable instruments currently available on medium flux reactors.

# CANS designs

## The ESS-B reference design
The ESS-Bilbao institute is in charge of the Spanish contribution to the ESS construction. It has put together a detailed technical design study of a CANS design which could provide neutrons as a user facility [20]. The reference design is based on a 50 MeV proton accelerator and a power on the target of 115 kW. It is using a rotating Beryllium target.

## The HBS reference design
The Jülich Center for Neutron Scattering at the Forschung Zentrum Jülich is considering the design of a High Brilliance Source with the following parameters, $E_p$ = 50MeV, $I_{peak}$ = 100mA, duty cycle 2%, P = 100kW, fixed Be target [11].

## The SONATE reference design
The CEA is considering a reference design SONATE with the following parameters, $E_p$ = 20MeV, $I_{peak}$ = 100mA, duty cycle = 4%, P = 80kW, fixed Be target. These parameters were chosen partly because they correspond to the first 20m of the ESS Linac (out of 600m). Hence the components (Source, RFQ and DTL) are available with no R&D developments.

## The LENOS design
The LNL Laboratori Nazionali di Legnaro is considering the LENOS design (LEgnaro NeutrOn Source facility). The design parameters are $E_p$ = 70 MeV, $I_{av}$ = 750 µA, P = 52.5 kW, Lithium target. This facility is close to completion but is not oriented towards neutron scattering but rather towards nuclear physics.

## The NOVA-ERA reference design
The Jülich Center for Neutron Scattering at the Forschung Zentrum Jülich is also considering the construction of a "laboratory" source with modest performances, NOVA-ERA Neutrons Obtained Via Accelerator for Education and Research Activities. The design parameters are $E_p$ = 10MeV, $I_{peak}$ = 1mA, P = 1kW, Be target, duty cycle 4-10%. Such a source can be built using off-the-shelf commercial proton accelerators.

# Neutron scattering instruments performances

Several institutes performed flux calculation at the sample position for several neutron scattering techniques and for several CANS designs (see Tables below presented during the workshop). In these simulations no constrains were set regarding the operation frequency, pulse length, instrument length, moderator material. These parameters were optimized according to the applications.

|  | Backscattering | Cold ToF | Thermal ToF |
|---|---|---|---|
| $E_{i,f}$ (meV) | 1.84 | 5 | 45 |
| $\frac{\Delta E_i}{E_i}$ (%) | 1 | 2 | 5 |
| $\Delta\theta(°)$ | 4 | 2 | 0.75 |
| $\Delta t(\mu s)$ | 120 | 50 | 18 |
| Rep. rate (Hz) | 200 | 100 | 400 |
| Flux (cm$^{-2}$s$^{-1}$) | $2.5 \times 10^7$ | $1.3 \times 10^5$ | $1 \times 10^5$ |
| Reference instrument | OSIRIS | LET | MERLIN |
| Flux reference (cm$^{-2}$s$^{-1}$) | $2.7 \times 10^7$ | $5 \times 10^4$ | $6 \times 10^4$ |

*Figure 8: Various Inelastic instruments performances using the HBS reference design. A comparison with equivalent instruments at ISIS is provided (J. Voigt et al, [21])*

|  | Reflectometer | SANS | Powder |
|---|---|---|---|
| F (Hz) | 48 | 48 | 96 |
| Dt (µs) |  |  | 40 |
| Moderator | Para-H2 | CH4 | thermal |
| Flux (n/cm²/s) | 1.3x10$^8$ n/cm²/s at 3mrad div | 2.4x10$^7$ n/cm²/s at 7% resolution | 6x10$^6$ n/cm²/s at 10mrad div |

*Figure 9: Various elastic instruments performances using the HBS reference design (U. Rücker, 2017).*

|  | Reflectometer | SANS | Powder | Imaging |
|---|---|---|---|---|
| F (Hz) | 48 | 48 | 96 |  |
| Dt (µs) |  |  | 40 |  |
| Moderator | Para-H2 | CH4 | thermal |  |
| Flux (n/cm²/s) | 5x10$^4$ n/cm²/s | 7x10$^4$ n/cm²/s at 10% resolution | 4.3x10$^3$ n/cm²/s at 0.3% resolution | 2.5x10$^3$ n/cm²/s at L/D=200 |

*Figure 10: Various elastic instruments performances using the NOVA-ERA reference design (U. Rücker, 2017).*

| Technique | Flux on sample | Reference spectrometers |
|---|---|---|
| **Reflectivity** | 0.8x10$^7$ n/s/cm² | HERMES@LLB 1x10$^7$ n/s/cm² <br> POLREF@ISIS ~1x10$^7$ n/s/cm² |
| **SANS** | 0.7x10$^6$ n/s/cm² (low Q) <br> 2.2x10$^6$ n/s/cm² (med Q) <br> 6.7x10$^6$ n/s/cm² (high Q) | PAXE@LLB (low Q) 0.7x10$^6$ n/s/cm² <br> SANS2D@ISIS 1x10$^6$ n/s/cm² |
| **Low resolution powder diffraction** | 2x10$^6$ n/s/cm² | G41@LLB 2x10$^6$ n/s/cm² |
| **Imaging (white beam)** | 1.5x10$^6$ n/s/cm² (for L/D = 240) <br> 1.3x10$^7$ n/s/cm² (for L/D = 80) | ICON@PSI 1x10$^7$ n/s/cm² <br> CONRAD@HZB 1x10$^7$ n/s/cm² (for L/D = 240) |
| **Imaging (time resolved)** | 1x10$^5$ n/s/cm² <br> (for L/D = 500) dl/l = 1% | ANTARES@FRM2 5x10$^5$ n/s/cm² |
| **Spin-Echo** | 2x10$^6$ n/s/cm² | MUSES@LLB 2x10$^7$ n/s/cm² (at 5A°) |
| **TOF** | 6x10$^7$ n/s/cm² | OSIRIS@ISIS 3x10$^7$ n/cm²/s |

*Figure 11: Various elastic instruments performances using the SONATE reference design (F. Ott, 2017).*

The general outcome of these calculations is that a high end CANS should make it possible to build instruments on par in terms of performances compared to existing medium flux reactors such as Orphée or spallation sources such as ISIS.

# Optimized operation conditions

In the case of low resolution instruments such as SANS, reflectometers or imaging, the obvious choice is to aim for a long pulse source which can be fully exploited. This is the same philosophy as applied at ESS. There is no consensus between very broad band instruments or shorter bandwidth instruments. Designs using a low frequency (20Hz) and broadband (→ 16A°) as well as designs using higher repetition rates (48Hz) and shorter bands (→8A°) have been considered. Very broad band instruments may be useful only for some scientific topics. The discussion is reminiscent of the ESS situation where two SANS machines are being built namely LoKI and Skadi which have slightly different designs. In the case of reflectometry where a broadband instrument is easy to exploit, a $CH_4$ moderator should be considered since it is providing more flux at long wavelengths. For SANS, a para-$H_2$ moderator should be used as it provides more flux.

For diffraction, the operation parameters are not straightforward. One could consider building long instruments on a low repetition rate source as is done at ESS (f=40 Hz, w=250µs, L = 50m) or build shorter instruments using pulse shaping choppers to achieve useful energy resolution (f=300 Hz, w=20µs, L = 15m). One should also consider bi-spectral extraction (see Helmholtz Zentrum Berlin for example). If short wavelength neutrons (~0.5A°) are available they could be used for diffraction or PDF measurements (e.g. SANDALS@ISIS).

In the case of spectroscopic measurements, the motto is "fill all the available phase-space". Operating at fast repetition rates (up to 400 Hz) allows to achieve large gains compared to existing spallation sources. The instrument design is simple provided you tune the source repetition rate to fill the phase-space. It also saves a number of choppers. This type of operation avoids using Repetition Rate Multiplication schemes as is necessary on CSPEC@ESS for example to fully exploit the ESS available intensity. Nevertheless, the details of the neutron pulse shape should not be included as a design requirement and pulse shaping choppers should be considered for all instruments so that the neutron flux from the source can be optimized irrespectively from the scattering instruments resolution. Pulse shaping choppers also offer extra flexibility in the instrument resolution choices by the user.

Spin-Echo is non trivial to implement efficiently on a pulsed source. It is also one of the instrument whose performances pales on a CANS compared to an equivalent instrument on a reactor. It might benefit from the possibility to achieve very large divergence out of a CANS moderator but this would require to operate resonant Spin-Echo which seems most adapted to benefit from large divergences. The development of suitable correction coils remains an open issue.

The possibility to produce Very Cold Neutrons (VCN) might mostly benefit to Spin-Echo techniques but there is currently no return of experience in the moderation at temperatures below 20K. The cross sections at these very low temperature are unknown which make any calculation unreliable. The use of very cold neutrons is not trivial for most classical neutron scattering techniques. In the case of SANS, one should for example take care of multiple scattering. In any case, an experimental program should be endeavored to get experimental data before considering instrumentation using VCN.

As a rule, the source should be built "around' the instruments. A scientific goal should be clearly defined and a source accelerator should be built accordingly to fulfill the requirements. This view is quite different from the current process, where the biggest possible source is built and then instruments are fitted as best as possible.

## What are the instruments the most suited on a CANS ?

CANS will never provide performances on par with high-end spallation sources such as ESS. Hence simple workhorse instruments not flux limited should be considered: SANS, powder diffraction, reflectometer, and radiography. Such instruments are technically "low risk". They are also generic so that they are not prone to obsolescence (over 20-30 years) and they are also high output instruments.

In parallel with low risks instruments, the relatively easy operation of a Compact Neutron Source opens possibility for Instrumental Developments (e.g. as performed at TU Delft). It will for example be unrealistic to "try" new instrumental concepts at ESS due to cost and regulation issues. Small sources are a lot more adapted.

It might also be significantly easier to "invite" University users to perform specific long term developments around the source without aiming for ultimate performances but for "novelty". The instrument would thus not be optimized in terms of raw performances but in terms of capability (for example very high pressures or very high temperatures). The measuring times would not be a design parameter.

## Possible new instruments concepts

In the case of SANS, setups using multi-beam collimation could be considered (e.g. NIST, 7 beams).

In the case of reflectivity, a focusing reflectometer of the ESTIA type could boost the efficiency of the instrument by a factor 10 (for small samples). While such an instrument would have performances far better than current instruments its cost would also be significantly higher than a plain reflectometer. Return of experience should be gathered on AMOR II@PSI.

In the field of radiography, a strong limitation is given by the detection efficiency of the (scintillator + camera) setups. New detection systems using Micro channel plates are becoming commercially available (e.g. Proxivision). While the price of these systems is higher than camera setup (k€100) the gain in detection efficiency (x5 to x10) fully justify their purchase.

The use of transmission diffraction on pulsed sources has been recently investigated by Mimiya et al [22]. It is suggested that the technique may easily be implemented on CANS sources in an efficient way for characterization purposes.

CANS also offer the opportunity to start neutron optical systems very close to the moderator. Hence the phase space which can be transported from the source to the sample can be significantly increased. However, there are currently very few techniques which may benefit from this possibility since highly divergent beams are more difficult to use efficiently. Innovative spectrometer designs should be investigated.

Less traditional applications may be considered such as fast neutron imaging for engineering applications.

PGAA using the pulsed structure opens perspectives for new types of measurements. Such aspects need to be further investigated.

The field of very cold neutrons for neutron scattering is still virgin. Some work program around this topic could be considered, starting with the acquisition of nuclear data on the moderation processes at very low temperatures.

## Moderators

The Target-Moderator-Reflector (TMR) system will be the key component of a CANS. The instrument suite planned on a CANS will be used to define the key parameters of the TMR system.

Current work on the TMR design has led to the following conclusions. The cost of a target station is expected to be low (~0.5-1M€). Each instrument should have its own moderator (included as part of the instrument, e.g. MACS@NIST). The cost of a moderator is expected to be low though (~100k€). A single TMR assembly can serve up to 5 instruments with marginal flux penalty (<10%). This strongly supports avoiding sharing beam ports. Note however that a wedge opening in the moderator can serve several instruments without flux penalty (this might be used for example for SANS or reflectometry). A single target station can efficiently accommodate thermal and cold sources with marginal flux penalty (10%). This makes the bispectral extraction even more appealing.

The moderator should not be optimized for a specific pulse shape. Pulse shaping should be achieved with mechanical devices at the exit of the moderator. Note that the limited requirements in terms of shielding make it possible to setup the choppers very close from the source.

## Optics

Neutron optics has made tremendous progress during the last 25 years. It may sound appealing to benefit from these progress in the instrument designs and implement advanced optics as early as possible inside the moderator. Unfortunately few instruments can benefit from large angular divergence so that a simple hole coupled to a straight or elliptic guide does the job. For diffraction an elliptic guide can be useful but only requires low m due to the low required divergence. Only inelastic instruments may use beam with high divergence (a few degrees). Note that brilliance transfer of thermal beams is not trivial though feasible.

The instruments on a CANS are generally rather short so that the guides are short and elliptic guides are not too big. On the other hand it might not be obvious to implement curved guides to get rid of fast neutrons. The direct line-of-sight might be avoided by using a bender inside the moderator. Filters such as sapphire or beryllium could also be considered at the moderator exit to reduce the fast neutron background.

## Shielding and background

Shielding issues are unclear since there is very little experience of CANS operation at high power. It is likely that the tool developed for ESS will be made available. In any case background noise simulations are non-trivial and the best route to follow is to perform experiments and gather experience.

One should note however, that the operation in pulsed mode allows reducing the background down to very low levels compared to continuous sources.

## Very cold neutrons

Since the design and construction of a TMR on a CANS is not too demanding, and also since the actual radiative heating on a CANS is very low (1-2mW) it might be appealing to consider the production of very cold neutrons using a VCN moderator. This opportunity was considered during a dedicated workshop was organized in Argonne in 2005 (Workshop on Applications of a Very Cold Neutron Source, Argonne, 2005) available on the Web [23].

Unfortunately very little is known about the cross sections at very low temperatures so that calculations are rather unreliable. Hence only experiments can provide data. Besides, due to the lack of availability of very cold neutrons very little effort has been made in developing VCN instrument designs. Spin-Echo is seen as the technique with most potential to benefit from very cold neutrons as the gains scale as $\lambda^4$. One should beware of multiple scattering though.

Reflectometry using refractive devices (RAINBOW type would be significantly easier to implement (higher refraction angles, better detection efficiency).

## Benchmarking

Benchmarking of instruments simulations codes (McSTAS – Vitess) is well established. Some benchmarking should be done for Monte-Carlo calculations using MCNP Vs GEANT4 Such comparisons have shown that the output are not perfectly equivalent, sometimes because of bugs, sometimes for unidentified reasons. Eventually Monte-Carlo simulations should be compared with experimental data. Note that CANS are a perfect case where this is easy to perform. The LENS source has indeed been used to validate concepts for SNS. SNS is also currently building a CANS for validation purposes for the Target Station 2 project.

There is currently very little possibility of interfacing various Monte-Carlo simulations codes such as MCNP – GEANT4 – McStas – Vitess. Efforts are under way. The MCPL format has been proposed to exchange data between various codes. Comblayer aims at providing a higher level interface on top of MCNP.

## Other actions

It would be profitable for the neutron community to broaden the number of people interested in CANS especially in the field of simulations. This is difficult at the moment due to the ESS developments but it is also an opportunity since a lot of people are thinking about instrument design and have off-the-shelf TOF instrument designs.

We could consider providing "guidelines" to people interested in designing an instrument from scratch: (i) Source components + time structures for McStas or Vitess, (ii) thermal and cold spectra, (iii) operation possibilities, frequency range (10-1000Hz), pulse length range (100-1000µs).

## Conclusion

It seem very likely that high end CANS are a way to provide neutron to high performance neutron scattering instruments in a cost-effective way. The designs and performances of the sources are rather well bound even though some technical challenges remain especially with respect to the target.

Besides, the source design can be flexible and one may aim for lower flux sources (not too expensive) but sufficient to fulfill the needs of a University or a large institute, or aim for a high-end CANS whose objective is to replace medium flux nuclear reactor user facilities.

The simulation shows that the operation parameter optimization strongly affects the performances of the neutron scattering instruments. The question arise if it would make sense to have a source optimized for low resolution instruments (long pulse, low repetition rate) in parallel with sources optimized for higher resolution instruments (short pulses, high repetition rates). Combining these two types of operation might not be trivial. A network of European sources with each a domain of excellence could be considered. Specifically in Europe, CANS may be a way of performing high quality science. Smaller national sources are an essential component to maintain an active and competent user base which is an essential component of a successful ESS in the future.

Also the question arises regarding the positioning of CANS with respect to research reactors. In Europe it is rather unlikely that new nuclear research reactors will be built. Hence CANS appear as a possible replacement for these reactors and a way to maintain the European expertise in neutron scattering. In countries outside Europe, the motivation to use nuclear research reactors as a first step to build "nuclear" expertise is still vivid. This may change in the medium term. Note that the projected cost of CANS is significantly lower than reactors and spallation sources so that it lowers the barrier into neutron scattering (or other neutron science). Hence CANS might be a fruitful way to open new opportunities for the use of neutrons.

CANS will be less a subject to complicated administrative rules since they are usually not considered as nuclear facilities. The operation cost will be lower than on current facilities and hence the operation might be more flexible. One may more easily consider to dedicate a large part of the beam time to "fast access" which is often requested both in material sciences as well as by industrial users. Long term instrumental programs can easily be considered due to less constraints compared to a reactor. Also a lower cost opens the possibility to initiate long term programs with universities, also on dedicated instruments.

Besides the potential of providing high performance neutron instrumentation, one of the most appealing aspect of CANS is to potentially turn neutrons into "laboratory" particles. Compared to X-rays where a very wide range of facilities are available from table-top diffractometers to X-FEL labs. Neutrons have until now lacked "easy" access for users being restricted to large scale facilities. While ESS will be the brightest neutron source in the world, it will provide only a small fraction of the current instrument-days (4000 out of 35000) so that by 2050 neutron scattering may turn to a niche technique with a very small user base which will make it impossible to justify going beyond ESS., On the other hand, a network of CANS may open a number of possibilities not foreseen yet in the use of neutrons, not only in scattering to make it a routine tool.

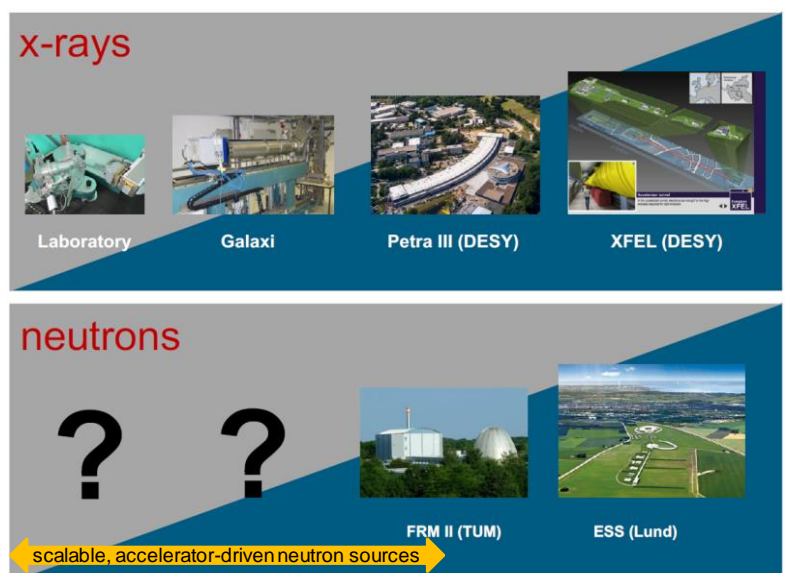

*Figure 12: CANS have the potential to eventually propose a consistent neutron source panorama so as to fullfil all potential users needs (Adapted from Thomas Brückel, FZ Jülich).*

# Annexe 1: Convention to be adopted for Monte-Carlo calculations

It appears that there is no consensus about fast, epithermal, thermal, cold neutron range so that Monte-Carlo simulations providing neutron flux in some energy range can hardly be compared.

We propose that after Monte-Carlo simulations, the binnings follow this convention. This will allow to compare calculation performed at different places using different codes.

| | | |
|---|---|---|
| Fast | E>10keV (anything that cannot be used) | |
| Resonances neutrons | 2eV < E < 10keV | |
| Epithermal Neutrons | 500meV < E < 2eV | |
| Thermal | 10 meV < E < 500meV | 0.41 A° - 2.9 A° |
| Cold | 1meV < E < 10meV | 2.9 – 9.2 A° |
| Very Cold | E < 1meV | 9.2 A° |

# Annexe 2: Instrument classification

This table aims at providing typical parameters for a broad range of instruments

| | Beam size A | Divergence | Bandwidth dλ/λ |
|---|---|---|---|
| SANS | 10x10mm² | 0.3° | 0.1 |
| Reflectivity | 40x1mm² | 5° x 0.05° | 0.1 |
| Radiography | 100x100 mm² | 0.15° | 1 |
| PGAA | 30x30mm² | 5° | 1 |
| Powder diffraction | 10x10mm² | 0.6° | 0.01 |
| Single crystal diffraction | 5x5mm² | 0.6° | 0.01 |
| Direct TOF cold | 30x30mm² | 3° | 0.03 |
| Direct TOF thermal | 30x30mm² | 3° | 0.03 |
| TAS | 30x30mm² | 5° | 0.04 |
| Indirect TOF | 30x30mm² | 3° | 0.03 |
| Spin-Echo | 30x30mm² | 5° | 0.2 |
| Backscattering | 30x30mm² | 5° | |
| Time focussing | 30x30mm² | 3° | 0.1 |

## Annex 3: Instrument benchmark

The following table aims at compiling "reference" instruments well known by users and rather well documented in terms of performances. These instruments should be used as benchmarks for instruments designed to operate on a CANS.

| Technique | Instrument | Reference |
|---|---|---|
| SANS | SANS1@MLZ | S. Mühlbauer, NIM A 832 (2016) 297 |
|  | KWS1@MLZ | H. Frielinghaus et al., J. Large-scale Res. Facilities, 1, A28 (2015) |
|  | SANS2D@ISIS | ISIS website |
| Reflectivity | EROS@LLB | F. Cousin, EPJ+ 126 (2011) 109 |
|  | FIGARO@ILL | R.A. Campbell et al, EPJ+ 126 (2011) 107 |
| Radiography | ICON@PSI | A. Kaestner, NIM A 659 (2011) 387 |
|  | ANTARES@MLZ | U. Garbe, Physics Procedia 69 (2015) 27 |
|  | IMAT@ISIS | T. Minniti et al, NIM A 888 (2018) 184 |
| Powder diffraction | G41@LLB | ? |
| Single crystal diffraction |  |  |
| PGGA | KFKI | ? |
|  | MLZ | Z. Revay et al., J. Large-scale Res. Facilities, 1, A20 (2015) |
| Direct TOF | IN5@ILL | J. Ollivier, Physica B 350 (2004) 173 |
|  | LET@ISIS | R.I. Bewley, NIM A 637 (2011) 128 |
|  | MERLIN@ISIS | R.I. Bewley et al, Physica B Cond.-Matt.385-86 (2006) 1029 |
| Indirect TOF | OSIRIS@ISIS |  |
| Spin-Echo | MUSES@LLB |  |
| Backscattering | SPHERES@MLZ | M. Zamponi et al., J. Large-scale Res. Facilities, 1, A30 (2015) |
| TAS | BiFROST@ESS |  |

## References

[1] IAEA database; nucleus.iaea.org/RRDB/RR/ReactorSearch.aspx
[2] IPSN Intense Pulsed Neutron Source: www.aai.anl.gov/history/project_pages/ipns.html
[3] LANSCE: lansce.lanl.gov/
[4] ISIS: www.isis.stfc.ac.uk/
[5] SNS: neutrons.ornl.gov/sns
[6] JPARC: j-parc.jp/index-e.html
[7] ESS: europeanspallationsource.se/
[8] ESFRI Report, *Neutron scattering facilities in Europe, Present status and future perspectives.*
[9] C.M. Lavelle et al, NIM A **587** (2008) 324-341. *Neutronic design and measured performance of the Low Energy Neutron Source (LENS) target moderator reflector assembly.*
[10] S. Halfon, A. Arenshtam, D. Kijel, et al. Appl. Rad. Iso. **106** Special Issue (2015) 57-62.
[11] U. Rücker, T. Cronert, J. Voigt, J.P. Dabruck, P.-E. Doege, J. Ulrich, R. Nabbi, Y. Beßler, M. Butzek, M. Büscher, C. Lange, M. Klaus, T. Gutberlet, and T. Brückel, , Eur. Phys. J. Plus (2016) 131: 19. DOI 10.1140/epjp/i2016-16019-5. *The Jülich high-brilliance neutron source project.*
[12] X. Wang et al, Physics Procedia **60** (2014) 186-192. *Delivery of 3-MeV proton and neutron beams at CPHS: A status report on accelerator and neutron activities at Tsinghua University.*

[13] Y. Yamagata, K. Hirota, J. Ju et al. J Radioanal Nucl Chem **305** (3) (2015)   787-794. *Development of a neutron generating target for compact neutron sources using low energy proton beams.*
[14] F. Sordo et al, Nuclear Instruments and Methods in Physics Research A **707** (2013) 1–8. N*eutronic design for ESS-Bilbao neutron source.*
[15] www.indiana.edu/~lens/sans.html
[16] UCANS, www.ucans.org/facilities.html
[17] R.N. Sinclair et al, NIM **117** (1974) 445-454
[18] N.C. Das, Langmuir 2012, 28, 11962−11968, dx.doi.org/10.1021/la2022598.
[19] M. Furusaka et al, Physics Procedia **60** (2014) 167-174. *Activity of Hokkaido University Neutron Source, HUNS*
[20] ESS-Bilbao, *Technical Design Report: ESS-BILBAO Target Station.*
[21] J. Voigt et al, Nuclear Inst. and Methods in Physics Research, A 884 (2018) 59–63.
[22] H. Mamiya et al, Scientific Reports | 7: 15516 | DOI:10.1038/s41598-017-15850-3
[23] B. J. Micklich and J. M. Carpenter, Proceedings of the Workshop on the Applications of a Very Cold Neutron Source, 21-24 Aug. 2005, Argonne National Lab., ANL-05-42 (2005).